# Nucleon-Nucleon Correlations and Inclusive Electron Scattering off Few-Nucleon Systems at $x > 1$

Silvano Simula *

Istituto Nazionale di Fisica Nucleare, Sezione Sanitá, Viale Regina Elena 299, I-00161 Roma, Italy

**Abstract.** Inclusive electron scattering off few-nucleon systems is investigated at $x > 1$ and high momentum transfer, including the contributions from quasi-elastic and deep inelastic scattering processes. It is shown that at $x > 1$ the inclusive cross section is dominated by the process of virtual photon absorption on a pair of correlated nucleons both in case of quasi-elastic and deep inelastic scattering events. The sensitivity of the nuclear response to the effects arising from the possible presence of multiquark cluster configurations at short inter-nucleon separations as well as from possible medium-dependent modifications of the nucleon structure function, is illustrated.

The aim of this contribution is to show that the investigation of inclusive electron scattering off few-nucleon systems at high values of $Q^2$ and $x > 1$ (where $Q^2$ is the squared four-momentum transfer and $x = Q^2/2M\nu$ the Bjorken scaling variable) could provide relevant information on the short-range properties of hadronic matter. According to the value of the invariant mass $W$ produced by virtual photon absorption on a nucleon in the nucleus, the inclusive process $A(e, e')X$ is governed by the following two mechanisms: i) the quasi-elastic ($QE$) process, for which $W = M$, and ii) the deep inelastic scattering ($DIS$) corresponding to $W > M$, where $M$ is the nucleon mass. The basic quantities that could be extracted from the analysis of $QE$ and $DIS$ processes at $x > 1$ are the nucleon momentum distribution in a nucleus and the valence quark distributions in a bound nucleon, respectively. However, present experimental data do not reach sufficiently high values of $Q^2$, so that at large values of $x$ ($> 1.3$) they are mainly due to $QE$ scattering and not to $DIS$ [1]. Nevertheless, the analysis of existing data (limited to $Q^2 \sim$ few $(GeV/c)^2$) shows that the kinematical regions corresponding to $x > 1.3$ are strongly affected by high momentum and high removal energy components of the nuclear wave function arising from nucleon-nucleon ($NN$) short-range and tensor correlations, provided that a consistent treatment of $NN$ correlations both in initial and final nuclear states is considered. Such a result has been obtained in [2], where it has been shown that in case of light as well as complex nuclei the inclusive cross section at $1.3 < x < 2$ is dominated by virtual photon absorption on a pair of correlated nucleons and by their elastic rescattering in the continuum.

---

* *E-mail address:* simula@hpteo2.iss.infn.it

As $Q^2$ increases, the effects due to the final state interactions in the $QE$ process are expected to decrease, whereas the $DIS$ contribution increases. In what follows, the inclusive process $A(e,e')X$ will be investigated at $x > 1$ and for values of $Q^2$ in the range $10 \div 20$ $(GeV/c)^2$, adopting the impulse approximation. The cross section for the inclusive process $A(e,e')X$ can be written as: $d^2\sigma/dE_{e'}d\Omega_{e'} = \sigma_{Mott} [W_2^A(x,Q^2) + 2W_1^A(x,Q^2)tg^2(\theta_e/2)]$, where $W_i^A(x,Q^2)$ ($i = 1,2$) is the inclusive nuclear response function, given within the impulse approximation by the following convolution formula

$$W_i^A(x,Q^2) = \sum_{N=1}^{A} \int dW dk dE \, \delta\left[\nu + k^0 - \sqrt{W^2 + (k+q)^2}\right]$$
$$P^N(k,E) \frac{M}{E_k} \frac{W}{\sqrt{W^2 + (k+q)^2}} \sum_{j=1,2} a_{ij} \, W_j^N(W,Q^2,...) \quad (1)$$

where $k \equiv |k|$, $k^0 \equiv M_A - \sqrt{(M_A + E - M)^2 + k^2}$ is the initial nucleon energy in the lab system (with $E$ being its removal energy) and the kinematical coefficients $a_{ij}$ are: $a_{11} = 1$, $a_{12} = k_\perp^2/2M^2$, $a_{21} = (Q^2/|q|^2)[1 - (Q^2/\bar{Q}^2)]$, $a_{22} = (E_k - k_\parallel \bar{\nu}/|q|)^2)/M^2 + (Q^2/|q|^2) \, k_\perp^2/2M^2$, with $\bar{Q}^2 \equiv |q|^2 - \bar{\nu}^2$, $\bar{\nu} \equiv \nu + k^0 - E_k$, $E_k \equiv \sqrt{M^2 + k^2}$, $k_\parallel \equiv k \cdot q/|q|$ and $k_\perp^2 \equiv k^2 - k_\parallel^2$. In Eq. (1) $W_i^N(W,Q^2,...)$ is the inclusive response function of a bound nucleon, which can be written as the sum of an elastic term and an inelastic one, viz.

$$W_i^N = W_i^{N(el)} + W_i^{N(inel)} \quad (2)$$

Following the $cc1$ prescription of Ref. [4], the elastic part reads as follows

$$W_1^{N(el)} = \frac{\bar{Q}^2}{4M^2} \{F_1^N(Q^2) + F_2^N(Q^2)\}^2 \, \delta(W - M)$$
$$W_2^{N(el)} = \{[F_1^N(Q^2)]^2 + \frac{\bar{Q}^2}{4M^2}[F_2^N(Q^2)]^2\} \, \delta(W - M) \quad (3)$$

where $F_{1(2)}^N(Q^2)$ are the Dirac and Pauli (free) nucleon form factors. As far as nucleon inelastic channels are concerned, the free nucleon structure function $W_i^{N(inel)}(W,Q^2)$, parametrized as in Ref. [3], will be adopted.

The relevant nuclear quantity appearing in Eq. (1) is the nucleon spectral function $P^N(k,E)$, which represents the joint probability to find in a nucleus a nucleon with momentum $k$ and removal energy $E$. In presence of ground-state $NN$ correlations, the nucleon spectral function can be written as $P^N(k,E) = P_0^N(k,E) + P_1^N(k,E)$, where $P_0^N$ includes the ground and one-hole states of the (A-1)-nucleon system and $P_1^N$ more complex configurations (mainly 1p-2h states) which arise from 2p-2h excitations generated in the target ground state by $NN$ correlations. For $A = 3$ and $A = 4$ one gets

$$P_0^N(k, E) = n_0(k) \, \delta(E - E_{min}) \qquad (4)$$

where $n_0(k)$ is the nucleon momentum distribution corresponding to the ground-to-ground transition and $E_{min} \simeq 5.5, \, 20 \, MeV$, respectively. As for the correlated part $P_1^N$, since high values of $k$ ($k > 1.5 \, fm^{-1}$) and $E$ ($E \sim (A-2)k^2/2(A-1)M$) are considered in this contribution, the extended two-nucleon correlation ($2NC$) model of Ref. [5] is adopted, viz.

$$P_1^{N_1}(k, E) = \sum_{N_2=n,p} \int dk_{CM} \, n_{rel}^{N_1 N_2}(k - \frac{k_{CM}}{2}) \, n_{CM}^{N_1 N_2}(k_{CM})$$

$$\delta[E - E_{thr}^{(2)} - \frac{A-2}{2M(A-1)}(k - \frac{A-1}{A-2} \, k_{CM})^2] \qquad (5)$$

where $n_{rel}^{N_1 N_2}$ ($n_{CM}^{N_1 N_2}$) is the momentum distribution of the relative (center-of-mass $(CM)$) motion of the two nucleons in a correlated pair, and $E_{thr}^{(2)} = M_{A-2} + 2M - M_A$ is the two-nucleon break-up threshold. Eq. (5) quantitatively reproduces the high momentum and high removal energy parts of the many-body spectral function calculated for $^3He$, $^4He$ and nuclear matter, as well as the momentum sum rule of complex nuclei (see for details Ref. [5]). In case of the deuteron one has $P_1^N = 0$ and $P_0^N(k, E) = n(k) \, \delta(E - E_{min})$, where $n(k)$ is the nucleon momentum distribution in the deuteron and $E_{min} = 2.226 \, MeV$ is the deuteron binding energy. It should be pointed out that a numerical investigation of the integration limits over $k$ and $E$ in Eq. (1) shows that the use of a non-relativistic spectral function is grounded for $x < 1.7$ (see also [5]).

Within the $2NC$ model nuclear effects in the inclusive process $^4He(e, e')X$ have been investigated, using the relative and $CM$ momentum distributions of Ref. [6]. The results obtained using in Eq. (1) the ground-to-ground spectral function $P_0^N$ only (Eq. (4)) and the full spectral function $P_0^N + P_1^N$, which includes the effects of $NN$ correlations (Eqs. (4-5)), are shown in Fig. 1. In order to illustrate the role played by binding effects, the results of the calculations performed by applying the closure approximation to the final nuclear states (i.e., by disregarding in Eq. (1) the $E$-dependence in the energy-conserving delta function), are reported in the same figure. It can be seen that at $x > 1$: i) the inclusive nuclear response is dominated by the process of virtual photon absorption on a pair of correlated nucleons both in case of $QE$ and $DIS$ processes; ii) binding corrections sharply affect the inclusive cross section, i.e. the knowledge of the momentum and removal energy behaviours of the nucleon spectral function is needed.

Let us now illustrate the sensitivity of the inclusive nuclear response to possible modifications of the short-range structure of the $NN$ cluster absorbing the virtual photon. To this end, let us consider the possibility that the nucleons in a correlated pair can loose their identity at short separations, so that the incoming photon can interact with a six-quark ($6q$) bag structure. This is the mechanism proposed and applied to the investigation of inclusive $DIS$ processes off the deuteron in Refs. [7]. For $A > 2$ the effects from the Fermi motion of the $CM$ of the $6q$ bag can be estimated by adopting the extended $2NC$ model, i.e. by using for the $CM$ momentum distribution of a $6q$ bag the one of a correlated $NN$ pair; this implies that the introduction of a $6q$ bag is assumed to modify only the intrinsic structure of a $NN$ cluster at short separations (see for details [8]). Furthermore, a medium-dependent modification of the structure function of a nucleon bound in a nucleus has been proposed in [9], based on the observation that small size components in the nucleon wave function interact weakly with the surrounding medium. Thus, the valence quark distribution should be suppressed in a bound nucleon with respect to the free nucleon case by a factor, which is expected to depend on the nucleon momentum $k$ [9]. The results of the calculations for the inclusive processes $^2H(e, e'X)$ and $^4He(e, e'X)$ are reported in Fig. 2 at $Q^2 \sim 15$ $(GeV/c)^2$. It can be seen that: i) at $x < 1.5$ the nuclear response could be sensitive to the medium-dependent modifications of the nucleon structure function, due to the point-like-configuration suppression of Ref. [9]); ii) the kinematical regions corresponding to $x > 1.5$ appear to be appropriate for investigating the effects of the possible presence of multiquark cluster configurations at short $NN$ separations; iii) the $QE$ contamination could be relevant at large values of $x$.

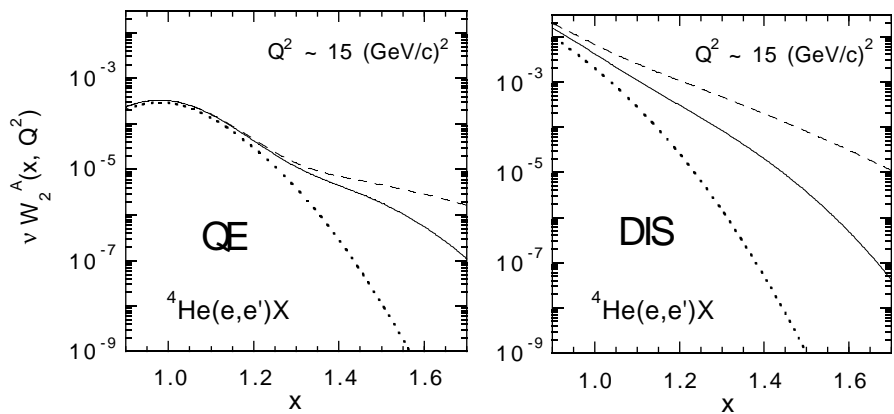

**Figure 1.** Nuclear structure function $\nu W_2^A(x, Q^2)$ for the inclusive process $^4He(e, e')X$ as a function of the Bjorken variable $x$ at $Q^2 \sim 15$ $(GeV/c)^2$. The dotted and solid lines correspond to the calculations of the $QE$ and $DIS$ contributions performed using in Eq. (1) the ground-to-ground spectral function $P_0^N$ only (Eq. (4)) and the full spectral function $P_0^N + P_1^N$ (Eqs. (4-5)), respectively. The dashed lines are the results obtained within the closure approximation for the final nuclear states.

In conclusion, inclusive electron scattering off few-nucleon systems has been investigated at $x > 1$ and high momentum transfer, showing that in such kinematical conditions the inclusive nuclear response is sensitive to the short-range structure of the two-nucleon cluster absorbing the virtual photon. It is therefore desirable that the investigation of inclusive $A(e,e')X$ processes at $x > 1$ will become part of the experimental program at existing and planned facilities (like, e.g., $CEBAF$, $HERA$ and $ELFE$).

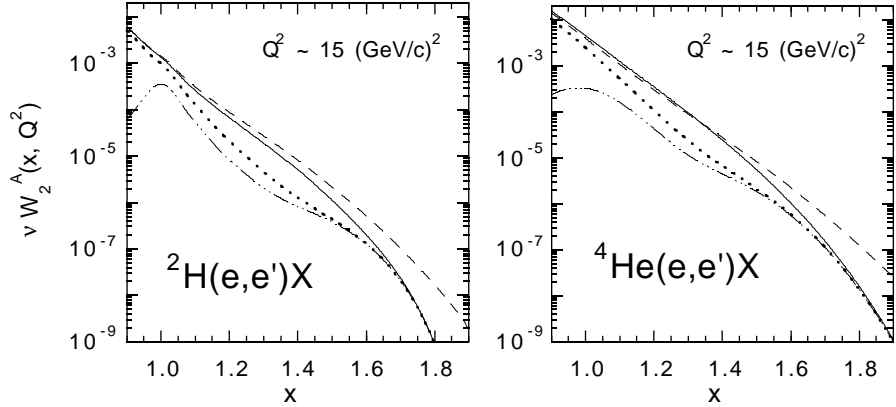

**Figure 2.** Nuclear structure function $\nu W_2^A(x, Q^2)$ for the inclusive processes $^2H(e,e')X$ and $^4He(e,e')X$ versus $x$ at $Q^2 \sim 15~(GeV/c)^2$. The dot-dashed lines are the $QE$ contribution, whereas the solid lines include the $DIS$ contribution calculated using in Eq. (2) the free nucleon structure function of Ref. [3]. The dotted lines correspond to the results obtained including in Eq. (2) the free nucleon structure function multiplied by the point-like-configuration suppression factor of Ref. [9]. Finally, the dashed lines are the results obtained assuming a 5% (10%) probability of a $6q$ cluster admixture in $^2H$ ($^4He$) ground-state.